\documentclass[12pt, a4paper, oneside]{book}

\usepackage[utf8]{inputenc}
\usepackage{setspace}
\usepackage{amsmath,amsfonts,amssymb,amscd,amsthm,xspace}
\usepackage{titlesec}
\usepackage{vmargin}
\usepackage{fancyhdr}
\usepackage{caption}
\usepackage{subcaption}
\usepackage{multirow}
\usepackage{multicol}
\usepackage{url}
\usepackage{tabularx}
\usepackage{graphicx}
\usepackage{epstopdf}
\usepackage{booktabs}
\usepackage{rotating}
\usepackage{listings}
\usepackage{float}
\usepackage[square, numbers, comma, sort&compress]{natbib} 
\usepackage[pdfpagemode={UseOutlines},bookmarks=true,bookmarksopen=true,bookmarksopenlevel=0,bookmarksnumbered=true,hypertexnames=false,colorlinks,linkcolor={black},citecolor={black},urlcolor={black},pdfstartview={FitV},unicode,breaklinks=true]{hyperref}
\hypersetup{urlcolor=black, colorlinks=true} 

\title{\thesisTitle}
\author{\authorName}
\date{\today}

\titleformat{\chapter}[display]
  {\normalfont\huge\bfseries\centering}
  {\chaptertitlename\ \thechapter}{18pt}{\Huge}

\setmarginsrb   { 3.0cm}  
                { 1.5cm}  
                { 2.0cm}  
                { 2.2cm}  
                { 0.3cm}  
                { 1.2cm}  
                { 0.3pt}  
                { 1.0cm}  

\begin{document}

\newcommand{\HRule}{\rule{\linewidth}{0.5mm}} 


\newcommand{\thesisTitle}{Single Channel Blind Dereverberation of Speech Signals}

\newcommand{\degree}{Master of Technology}

\newcommand{\authorName}{Dhruv Nigam}

\newcommand{\rollNo}{11D070032}

\newcommand{\dept}{Department of Electrical Engineering}

\newcommand{\college}{Indian Institute of Technology Bombay}

\newcommand{\currentyear}{2016}

\newcommand{\currentmonth}{July}

\newcommand{\supervisorOne}{Prof. V.Rajbabu}

\newcommand{\supervisorTwo}{}

\newcommand{\examinerOne}{}

\newcommand{\examinerTwo}{}

\newcommand{\chairman}{}

\frontmatter

\setstretch{1.5}


\newcommand{\titlePage}{

\thispagestyle{empty}
\begin{center}

\vspace*{15px}
{\Huge\bfseries \thesisTitle}\\[1.0cm] 

\textit{Submitted in partial fulfillment of the requirements\\[0.2cm] of the degree of}\\[0.2cm] \degree\\
in \\ 
Communication and Signal Processing \\[1.0cm]
\textit{by}\\[0.1cm]
\authorName \\[0.1cm] (\textit{Roll no.} \rollNo) \\[2.0cm]

\textit{Supervisor:}\\[0.2cm]
\supervisorOne \\[1.0cm]

\includegraphics[width=0.25\textwidth]{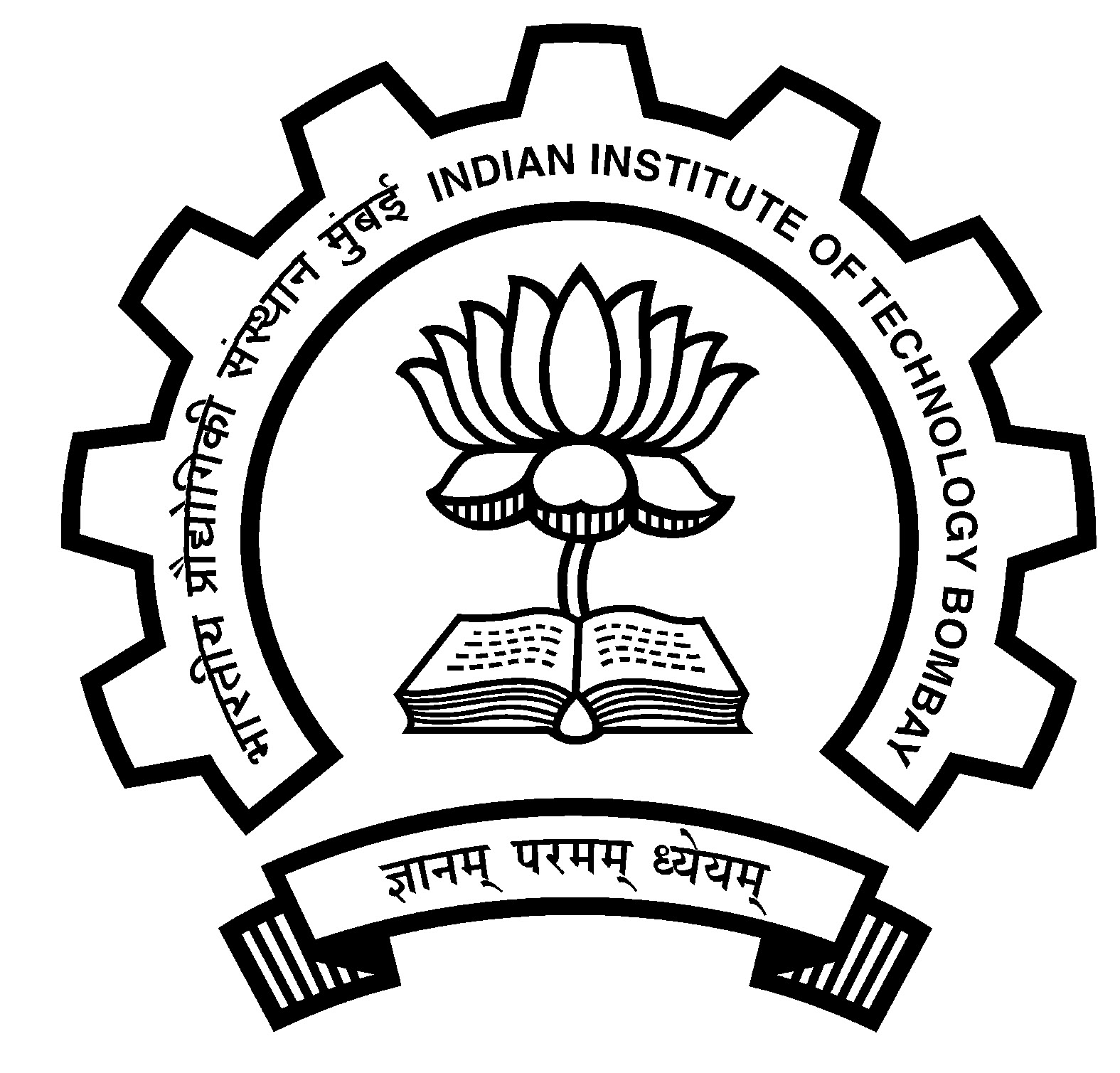}

\vspace*{10px}

\dept\\[0.2cm]
\college\\[0.2cm]

\currentmonth \ \currentyear\\ 

\end{center}

\clearpage
}


\newcommand{\approval}{
\thispagestyle{plain}

\begin{center}{\huge\bf M.Tech Dissertation Approval\par}\end{center}

\vspace*{15px}

\noindent This dissertation entitled 
\textbf{``\thesisTitle"}, submitted by 
\authorName  
 \ (Roll No.  \rollNo),
is approved for the award of degree of 
\degree \
in 
Communication and signal processing\\[1.0cm]

\begin{flushright}
\textbf{{Examiners}}\\[0.8cm]
\examinerOne\quad\rule{0.3\textwidth}{.3pt}\\[0.8cm]
\examinerTwo\quad\rule{0.3\textwidth}{.3pt}\\[1.6cm]

\textbf{{Supervisor}}\\[0.8cm]
\quad\rule{0.3\textwidth}{.3pt}\\[1.6cm]

\textbf{{Chairman}}\\[0.8cm]
\chairman\quad\rule{0.3\textwidth}{.3pt}\\[2.4cm]

\end{flushright}
\textbf{Date:} 05 \currentmonth { } \currentyear\\[0.3cm]
\textbf{Place:} \  Mumbai

\clearpage
}

\newcommand{\authorship}{
\thispagestyle{plain}

\begin{center}{\huge\bf Declaration of Authorship\par}\end{center}

\vspace*{15px}

\noindent I declare that this written submission represents my ideas in my own words and where others' ideas or words have been included, I have adequately cited and referenced the original sources.  I also declare that I have adhered to all principles of academic honesty and integrity and   have   not   misrepresented   or   fabricated   or   falsified   any   idea/data/fact/source   in   my submission.  I understand that any violation of the above will be cause for disciplinary action by the Institute and can also evoke  penal action from the sources which have thus not been properly cited or from whom proper permission has not been taken when needed.

\vspace*{10px}

\begin{flushright}
{Signature: ......................................\\[0.4cm]}

{\textbf{\authorName}\\[0.0cm]\rollNo\\[2.0cm]}

\end{flushright}
\begin{flushleft}
{Date: 05 \currentmonth { } \currentyear\\}
\end{flushleft}

\clearpage 
}

\newcommand{\abstractpage}{
\thispagestyle{plain}

\small {\noindent\authorName/ \supervisorOne{ }(Supervisor): \textbf{``\thesisTitle"}, \textit{M.Tech Dissertation}, \dept, \college, \currentmonth { } \currentyear.}\\[0.0cm]
\HRule\\[0.2cm]

\vspace*{10px}

\begin{center}{\huge{\textit {Abstract}}\par}\end{center}

\vspace*{10px}

\noindent Dereverberation of recorded speech signals is one of the most pertinent problems in speech processing. In the present work, the objective is to understand and implement dereverberation techniques that aim at enhancing the magnitude spectrogram of reverberant speech signals to remove the reverberant effects introduced. An approach to estimate clean speech spectrogram from the reverberant speech spectrogram is proposed. This is achieved through non-negative matrix factor deconvolution(NMFD). Further, this approach is extended using the NMF representation for speech magnitude spectrograms. To exploit temporal dependencies, a convolutive NMF based representation and a frame stacked model are incorporated into the  NMFD framework for speech.
A novel approach for dereverberation by applying NMFD to the activation matrix of the reverberated magnitude spectrogram is also proposed .
Finally, a comparative analysis of the performance of the listed techniques, using sentence recordings from the TIMIT database and recorded room impulse responses from the Reverb 2014 challenge are presented based on two key objective measures - PESQ and Cepstral Distortion.\\
Although, we were qualitatively able to verify the claims made in literature regarding these techniques, exact results could not be matched. The novel approach, as it is suggested, provides improvement in quantitative metrics, but is not consistent

\clearpage 
}

\titlePage

\setcounter{page}{1}



\addcontentsline{toc}{chapter}{Abstract}
\abstractpage

\pagestyle{fancy}

\lhead{\emph{Contents}} 

\tableofcontents 

\lhead{\emph{List of Figures}} 

\listoffigures 
\addcontentsline{toc}{chapter}{List of Figures}

\fancyhead{}
\rhead{\thepage}

\mainmatter


\chapter{Introduction} 


\ifpdf
    \graphicspath{{Chapter1/Figs/Raster/}{Chapter1/Figs/PDF/}{Chapter1/Figs/}}
\else
    \graphicspath{{Chapter1/Figs/Vector/}{Chapter1/Figs/}}
\fi

Speech processing and recognition have entered an exciting phase with voice-recognition application becoming ubiquitous across platforms. A prerequisite to capture any speech from the surroundings is the microphone. For many applications like video conferencing, and interactive home devices, the distance of the microphone from the speaker can be a few meters. Further processing or use of these speech signals, recorded from a distance is a challenge for two reasons. First, since the strength of the speech signal attenuates with distance, the background noise is more prominent in the recorded audio, than it would be for a close distance recording and proves obtrusive in recognition. Also, the speech signal may be reflected from multiple surfaces in the surroundings and cause an echo like effect in the recorded signal. This latter phenomenon is called Reverberation. Although, the human brain handles the effects of reverberation quite well, it severely distorts automatic speech recognizers’ capability to understand speech. \\

\section{Reverberation model}
\begin{figure}[h]
\centering
\includegraphics[scale=0.35]{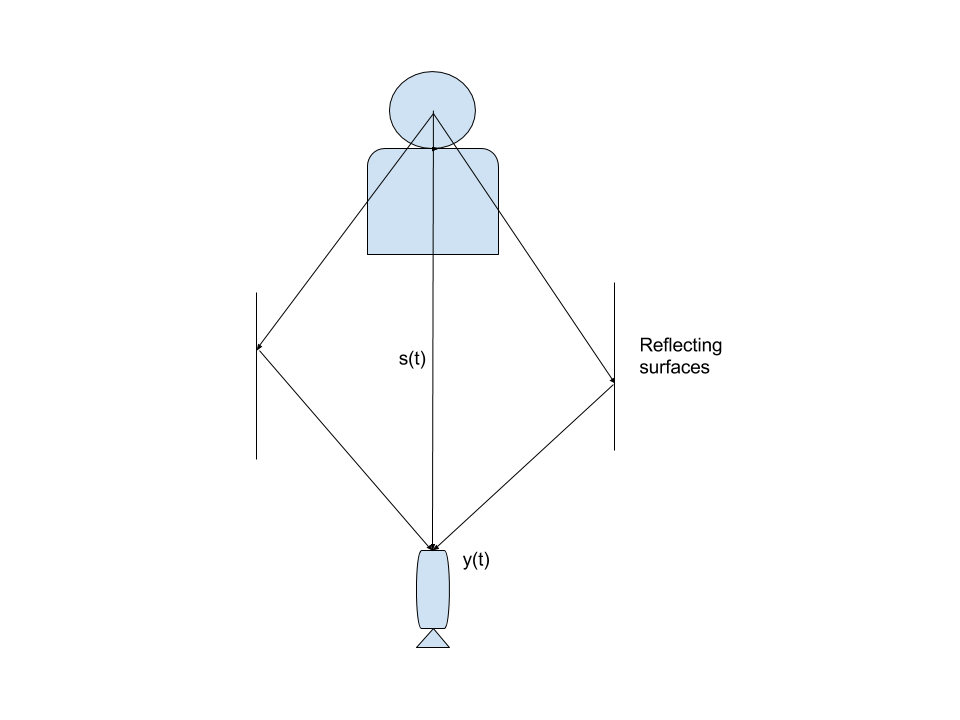} 
\caption{The reason for reverberation. Along with the source signal, several of its reflections from nearby surfaces are also recorded by the microphone}\label{Reverberation}
\end{figure}

Reverberation happens when an audio signal, after emanating from the source, undergoes multiple reflections from the surfaces in room. These reflections persist even when the source is switched off. The amount and nature of reverberation introduced in the speech signal depend on the surroundings in which the recording is done, as well as the relative positions of the microphone and the speaker. The degree of reverberation introduced by a particular enclosure or surrounding can be measured by the reverberation time (RT), also known as the T60. The Reverberation time is the time it takes for the signal strength to fall 60 db at the microphone, after the source is switched off. It is a characteristic property of the enclosure or environment where the source signal is recorded. Figure \ref{Reverberation} is an illustration of the mechanism of the process that leads to reverberation. In signal processing, the phenomenon of reverberation can be modeled as convolution of the source signal, $s(t)$ with a filter, $h(t)$.
\begin{align}
y(t) = s(t)*h(t) \label{eq :1}
\end{align}

Where $h(t)$ is the impulse response of the environment in which the source signal, $s(t)$ was generated and it depends on the relative positions of the microphone and the source and also the surroundings in which the signal is recorded. 

An example of a room impulse response is shown in Figure \ref{rir}. Each peak in the figure corresponds to a reflection arriving at the microphone after the original signal has arrived. Clearly there are a series of prominent reflections, temporally close to the direct signal(which arrived at time t=0), followed by weaker reflections which persist, at time up to 1000ms after the direct signal has died down. The reverberation time for typical office or home environments ranges from 200ms to 1000ms.  

\begin{figure}
\centering
\includegraphics[scale=0.7]{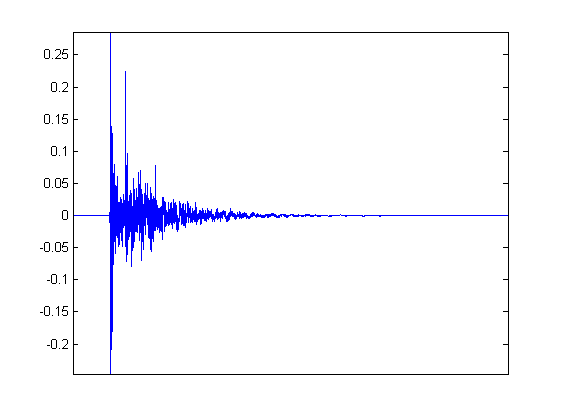} \label{rir}
\caption{An example of a typical time domain room impulse response. The strength of the reflections decays with time }\label{rir}
\end{figure}

\section{Automatic speech recognition and reverberation}
\begin{figure}
\centering
\includegraphics[scale=0.35]{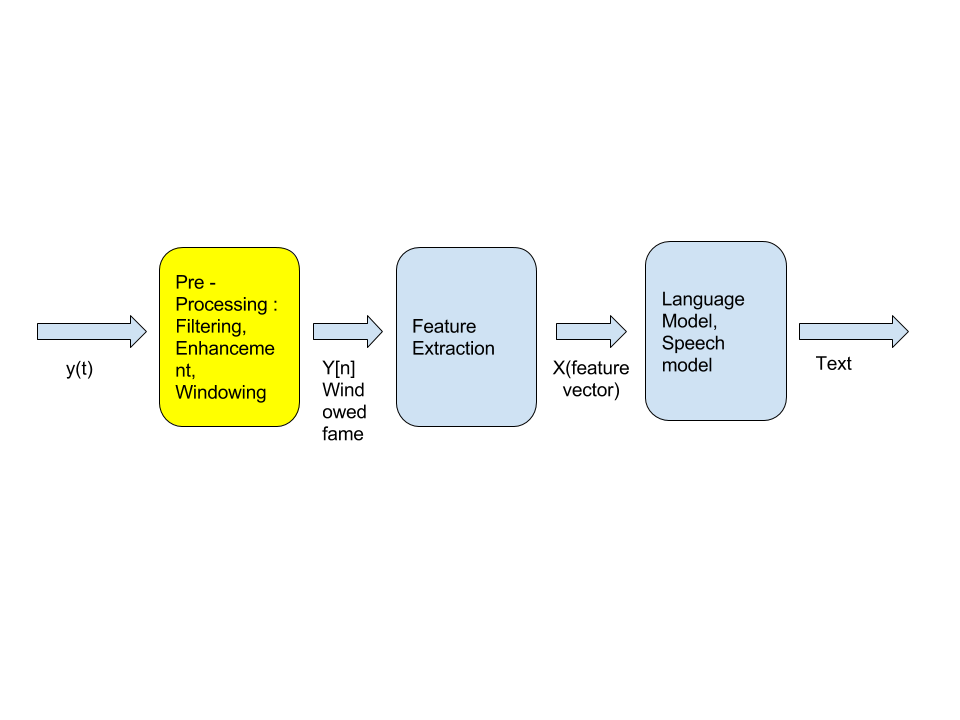} 
\caption{Schematic of an ASR system. All algorithms discussed here are applicable to the Pre-processing stage }\label{asr}
\end{figure}

The aim of this section is to give an overview of  an automatic recognition system. It will also describe the broad categories of dereverberation techniques. 
Figure \ref{asr}  shows the schematic of a typical automatic speech recognizer. The goal of the recognition process is to transcribe speech correctly into text. This whole process can be divided into two stages - the front end and the back end.

At the front end, necessary pre-processing is done to extract useful information for further processing and making the speech signal more compact by discarding irrelevant information and noise. The pre-processing stage attempts to make the incoming speech as independent of the room acoustics and the speaker, as possible for easy recognition. The common steps carried out at this stage are sampling, windowing, denoising and speech enhancement. 

The next critical step is feature extraction from the windowed and enhanced speech. Some information in the speech signal is not relevant to the recognition process. Thus, relevant features have to be extracted from the windowed signal. Furthermore, selecting a few features also reduces the dimensionality of the input data reducing computation time. Some prominent features such as the Mel-Cepstral features and wavelet domain features are used in almost all modern ASR applications. These features are extracted from windowed time frames of upto 40ms because of the quasi-stationary nature of speech in that time frame. That is, each segment of speech that is , say for instance, 30ms long is used to extract one feature vector.

After the incoming speech has been converted into a feature vector $X$, the ASR system has to estimate the sequence of words $w$, that would have given rise to $X$. It is here that acoustic and language models for speech are deployed. An acoustic model can, given a feature vector, predict the sound or phoneme that was uttered by the speaker and the language model provides the understanding of the language i.e. a probability distribution over sequences of words.
Mathematically, after the the signal $y(t)$ has been converted into a feature vector $x_{n \in T}$, where $T$ is the temporal the ASR system arrives at the correct transcription by solving the equation
\begin{align}
w^{*} = argmax_w p((X_n)_{n\in T} )p(w)
\end{align}
Here, p(w) gives the likelihood of observing $w$ in the present sequence of words and comes from the language model. $p((X_n)_{n\in T} )$, on the other hand quantifies the likelihood that $X$ would be observed, if $w$ was indeed spoken. This comes from the acoustic model.
\\

As mentioned earlier, the performance of the system depends on the quality of the features extracted. Typically, features are extracted from windowed segment of speech of length of 10-40ms because of quasi-stationary nature of speech in that duration. On the other hand, reverberation, as we have seen, can have deteriorating effects on the quality of speech, and thus the features extracted for upto 1000ms. This means that the impact of reverberation can last several frames. This poses a unique challenge and cannot be addressed like ordinary noise which is not correlated beyond a few ms.
\begin{figure}[H]
\centering
\begin{minipage}{.5\textwidth}
  \centering
  \includegraphics[width=.9\linewidth]{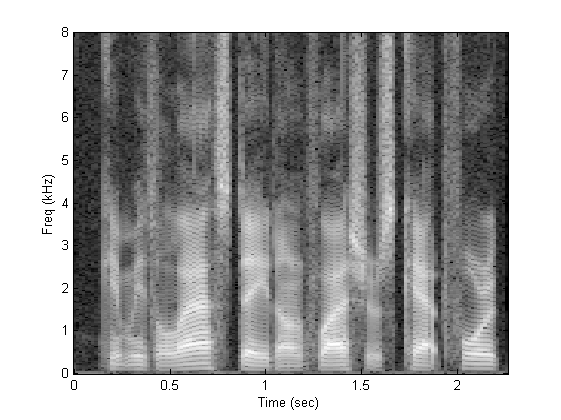} 
  \captionof{figure}{Magnitude spectrogram of a speech signal without any reverberation}\label{clean_ch1}
\end{minipage}%
\begin{minipage}{.5\textwidth}
  \centering
  \includegraphics[width=.9\linewidth]{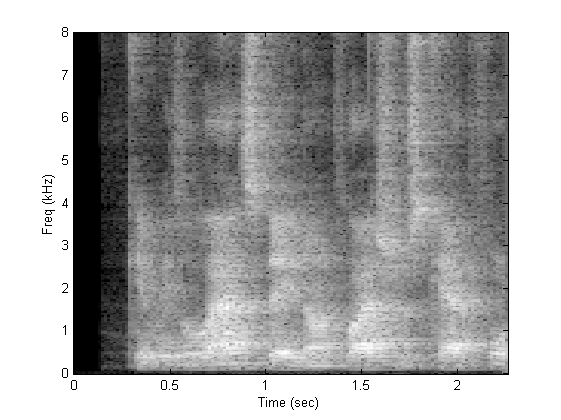} 
  \captionof{figure}{Magnitude spectrogram of the speech signal with reverberation.  }\label{input_normal_ch1}
\end{minipage}
\end{figure}

Figure\ref{clean_ch1} and \ref{input_normal_ch1} show the effect of reverberation on the magnitude spectrogram of a speech signal due to reverberation. Clearly, there is a ‘smearing’ effect, apparent in the magnitude domain as well as the feature domain. The reverberation caused by a single segment of speech can distort features for several subsequent frames, depending on the Reverberation Time. This distortion of features causes the ASR to function sub-optimally. Techniques to tackle noise cannot be directly deployed to tackle this since reverberation distortion is non-stationary as well as correlated with the real speech signal.

The ways to tackle reverberation can be separated into two broad approaches - front end and back end approaches. Front end approaches aim at cleaning the reverberation away from the feature vectors which can then be fed into the back-end of the speech recognition system, which works as it does for clean speech. Back-end techniques on the other hand aim at changing the acoustic model or change the decoding scheme for the feature vectors to account for reverberation. 

Here, the focus is only going to be on front end based approaches. Further, although the choice of features to be extracted is a critical component of the recognition process, that line of investigation is not explored here. Since majority of the features are derived from the magnitude-spectrum of the speech signal (eg. MFCC coefficients), only those techniques which propose to enhance and remove the effects of reverberation the magnitude spectrum have been explored. The phase information will be discarded in all the following investigations from the spectrogram. This has an added advantage of making the processing independent of speaker movement  since slight movements will only change the phase of the spectrogram.

Also, it will be hereon assumed that only a single microphone is used for the recording, and no prior information about the speaker, the surroundings or the microphone is known. Thus the exact problem that is going to be addressed is Single Channel Blind Dereverberation. 

However, much work has been done on modeling speech. There are a few universal properties that apply to all speech signals. We will put these properties to use to model clean speech better and thus improve dereverberation performance for some algorithms. Although, the room impulse response in much more difficult to model, a few simplifying assumptions will be made.

The next chapter, Chapter 2 will conceptually describe Non-Negative matrix factorization(NMF) and its application to speech processing. Chapter 3 discusses popular dereverberation techniques and how NMF factorization of speech magnitude spectrograms can improve the efficiency of these algorithms.

\chapter{NMF in speech processing}  

Non-Negative matrix decomposition or factorization is an algorithm where the objective is to approximate the given non-negative data matrix  $V \in \mathbb{R}^{\geq 0, M \times N}$ as a product of two non-negative matrices. the basis vector matrix,  $W \in \mathbb{R}^{\geq 0, M \times R}$ and the activation matrix $H \in \mathbb{R}^{\geq 0, N \times R}$.The parameter $R$ specifies the rank of the decomposition and depends on the complexity of the matrix $V$ that is being decomposed. NMF has been widely used in audio processing where it is applied on the magnitude spectrogram. The Magnitude spectrogram is inherently non-negative, and thus lends itself well to the NMF algorithm.
Since the problem of NMF does not have an analytical solution, a numerical approximation is generally used. Specifically, an objective function is chosen that quantifies the approximation error and using the gradient descent algorithm, minimum error is achieved. Thus, the NMF problem can be modeled as a constrained optimization problem, where the objective is to minimize the error of reconstruction of the original data matrix $V$. In this chapter, the Kullback-Leiber divergence has been used to measure the error since it has a special significance for speech spectrograms. It is defined as - 

\begin{align}
D = \left \| V \odot ln \left (  \frac{V}{W.H} \right ) -V +W.H \right \| \label{kldiv}
\end{align}
 $\left \| \cdot  \right \|_F$ in the Forbenious norm, $V$ is the data matrix being factorized and $W$ and $H$ are the factors.
This error has to minimized subject to the non-negativity constraints to achieve an accurate approximation. Thus, this is a contrained optimization problem. In \cite{lee} a multiplicativs update rules for $W$ and $H$ matrices that assures convergence to minimum error. They as given as
\begin{align}
H = H \odot \frac{W^T \cdot \frac{V}{W\cdot H} }{W^T\cdot 1} \label{H updates}
\end{align}
\begin{align}
W = W \odot \frac{\frac{V}{W\cdot H}\cdot H^T   }{ 1 \cdot H^T} \label{W updates}
\end{align}

$ \odot $ is the Hamdard product, all divisions are element wise and $1$ is a $M \times N$ matrix with all elements set to 1.

The steps to calculating the Non-Negative matrix factorization of a given data matrix, $V$ are given as -
\begin{itemize}
\item Randomly initialize the $W \in \mathbb{R}^{\geq 0, M \times R}$ and $H \in \mathbb{R}^{\geq 0, N \times R}$ with non negative values and with an appropriate value for the rank of the decomposition $R$
\item Apply multiplicative updates to the matrices as in Equations \ref{H updates} and \ref{W updates}  and calculate error at every iteration.
\item Stop after the error is lesser than a present threshold. Since the error converges to a finite positive value which will be unknown before the decomposition, and a relatively small tolerance level might not let the algorithm terminate,  it is advisable to use the number of iterations as a parameter to set the endpoint for the algorithm. 
\end{itemize}

\section{Illustration of NMF for audio}
In Figure 2.1, shows the spectogram of an instrumental piece of music, played on a piano, with 4 distinct notes, some played multiple  times. The duration of the audio is 4.7 seconds and has been sampled at $16kHz$. NMF was applied on ,$V$, the magnitude spectrogram of the audio signal, shown in Figure 2.1. Figure 2.2 shows the spectrogram of $\hat{V} \approx W.H$. The number of basis vectors ($R$) was chosen to be 5 and NMF was allowed 100 iterations.

Visually, it seems apparent that NMF has done a good job of representing the spectrogram with a limited number of basis. The reconstructed audio has no perceptible artifacts \cite{A}.  This is to be expected because the data has a simple structure in terms of the temporal patterns in the spectrogram. The structure is due to the fact that non-percussion instruments like the piano have a finite number of notes and each note has a unique fundamental frequency and predefined harmonics.These qualities do not extend to speech signals. This makes it difficult for the conventional NMF approach to do well in the speech scenario.

The two matrices, $W$ and $H$ have special significace when NMF is applied to the magnitude spectrogram. $W$ is the matrix whose columns are the basis set for $\hat{V}$. $H$ is a stack of set of row $R$ row vectors that constitute the weights corresponding to each basis vector across the time axis of the magnitude spectrogram. $W$ is called the basis vector matrix and $H$ is called the activation matrix, since it determines where on the time axis of the manitude spectrogram, a particular basis vector is 'activated'.

\begin{figure}
\centering
\includegraphics[scale=0.4]{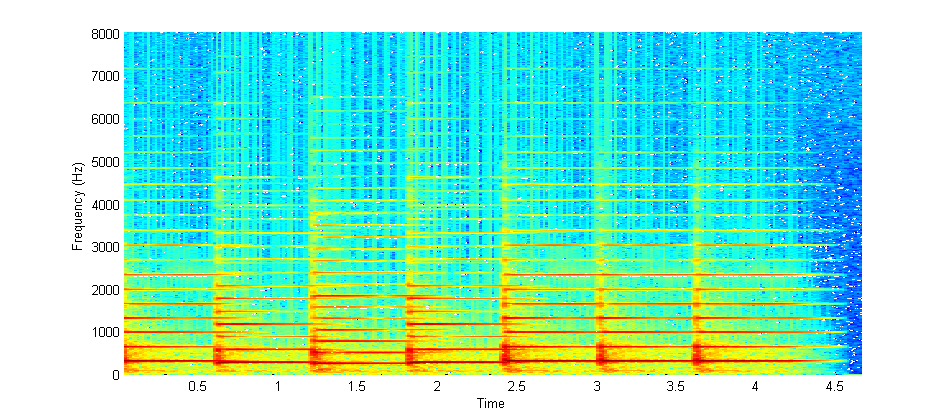}
\caption{Spectrogram of \emph{Mary}. Key STFT parameters - Sampling frequency : $16000kHz$. Window : hanning window of length 1024 samples. Overlap : 756 samples. FFT size 1024}
\end{figure}

\begin{figure}
\centering
\includegraphics[scale=0.4]{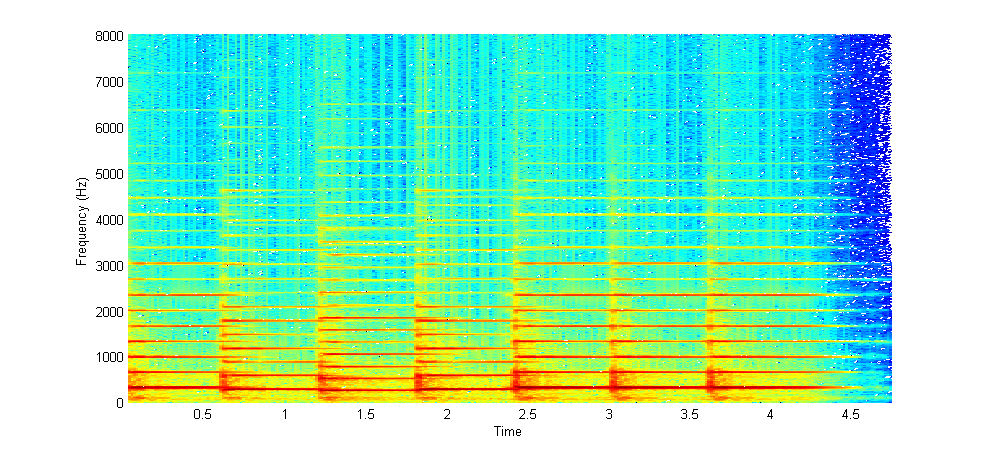}
\caption{Reconstructed Spectrogram of \emph{Mary}. Key STFT parameters - \textbf{Sampling frequency} : $16000kHz$. \textbf{Window} : hanning window of length 1024 samples. \textbf{Overlap} : 756 samples. \textbf{Size of fft}: 1024 }
\end{figure}

\begin{figure}
\centering
\begin{minipage}{.5\textwidth}
  \centering
  \includegraphics[width=.9\linewidth]{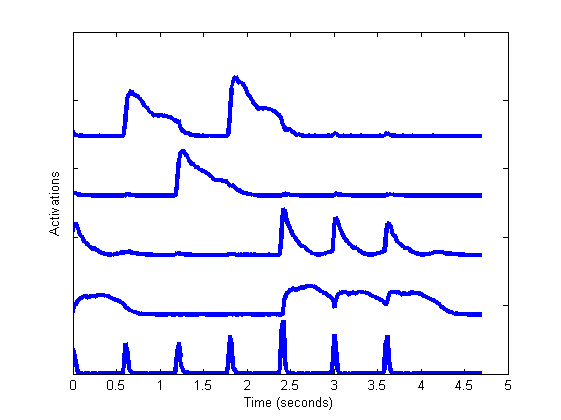}
  \captionof{figure}{The activation vectors corresponding to each basis vector}
  \label{fig:test1}
\end{minipage}%
\begin{minipage}{.5\textwidth}
  \centering
  \includegraphics[width=.9\linewidth]{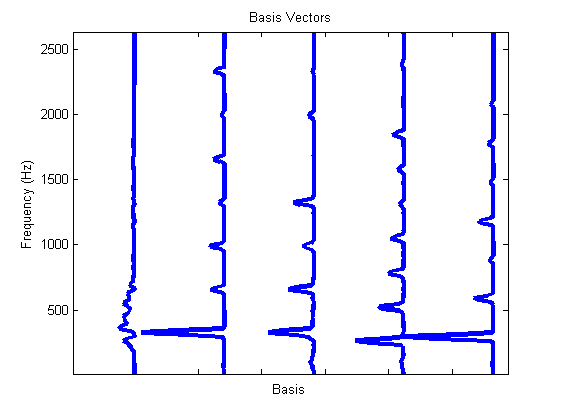}
  \captionof{figure}{Basis vectors. A total of 5 basis vectors were used. A zoomed version is shown to show the differences in frequencies}
  \label{fig:test2}
\end{minipage}
\end{figure}

Figure 2.3 and 2.4 show the activation times for each basis vector and the the basis vectors respectively. The activation vectors for the basis vectors from left to right are shown from bottom to top. It should be noted that the basis vectors are in fact magnitude spectra.\\

Refering to Figure 2.3 and 2.4, and analysing the activation times, we find that we can map them to the pattern of playing of a specific note at particular time instants. For example, the rightmost basis vector in the Figure 2.4 (corresponding to the topmost activation vector in Figure 2.3) gets activated at two time instants which are actually approximately the times of playing of a single specific note in the audio. It can also be seen that no other basis vector is getting activated during the period of activation of this basis vector. It follows that this basis vector has captured the expression of this note in the spectrogram. This is made clear when the audio reconstruction of the spectrogram achieved by setting all other activations to zero is heard \cite{A}. \\
Not all basis vectors though capture single notes this cleanly. Looking at the activations, it is also evident that the basis vectors corresponding to two activations (3rd and 4th from the top) get activated together whenever a particular note is played. To get the full expression of this note, we need both these basis vectors. \\
Also, the basis vector corresponding the the bottom most activation vector does not represent any single note, but is activated at the times of change of note. It can be inferred that this basis vector represents the complex part of the spectra when a note is just struck and the power is spread throughout the spectrum, just like in a percussion instrument.

\section{Convolutive NMF}
Figures 2.5-2.6 correspond to an audio sample, \emph{Machali}\cite{A}. It is an audio which is a speech recording of the sentence, \emph{"Machali jal ki hai rani, jeevan uska hai paani"}, in Hindi. It is 4.7 seconds long and is sampled at 16kHz. The figures  represent the magnitude spectrogram of the original audio, the reconstructed spectrogram via NMF using $R = 8$, the activation matrix and the basis vectors, in that order. Although the duration of the audio is the same as \emph{Mary}, to achieve the same accuracy, we need 8 basis vectors to represent the magnitude spectra. This is intuitive since speech signals have complex spectrogram patterns and thus have a higher rank than that of an instrumental piece, like \emph{Mary}.

\begin{figure}
\centering
\includegraphics[scale=0.6]{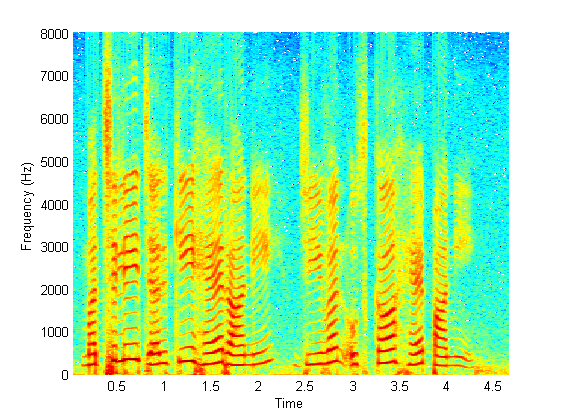}
\caption{Spectrogram of \emph{Machali}. Key STFT parameters - Sampling frequency : $16000kHz$. Window : Hanning window of length 1024 samples. Overlap : 756 samples. Size of fft: 1024 } \label{machali_spectrogram_ch2}
\end{figure}

\begin{figure}
\centering
\includegraphics[scale=0.6]{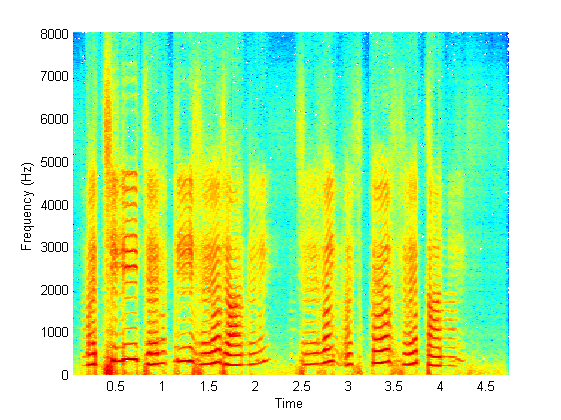}
\caption{Reconcstructed Spectrogram of \emph{Machali}. Key STFT parameters -, Sampling frequency : $16000kHz$. Window : Hanning window of length 1024 samples. Overlap : 756 samples. Size of fft: 1024 . R = 8 basis vectors were used to factorize the audio. Notice the almost identical magnitude spectrogram to that of the original signal spectrogram}
\end{figure}

\begin{figure}
\centering
\begin{minipage}{.5\textwidth}
  \centering
  \includegraphics[width=.9\linewidth]{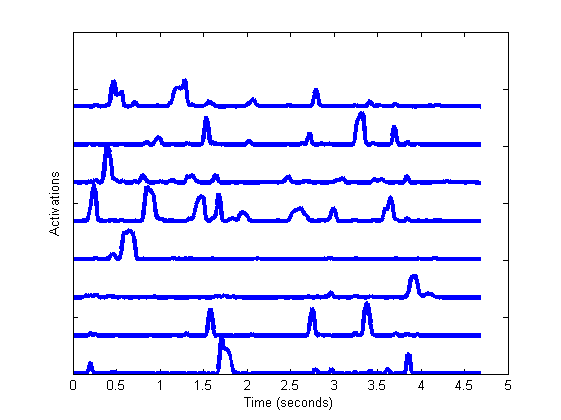}
  \captionof{figure}{Activation vectors for Machali}\label{activations_machali_nmf}
  \label{fig:test1}
\end{minipage}%
\begin{minipage}{.5\textwidth}
  \centering
  \includegraphics[width=.9\linewidth]{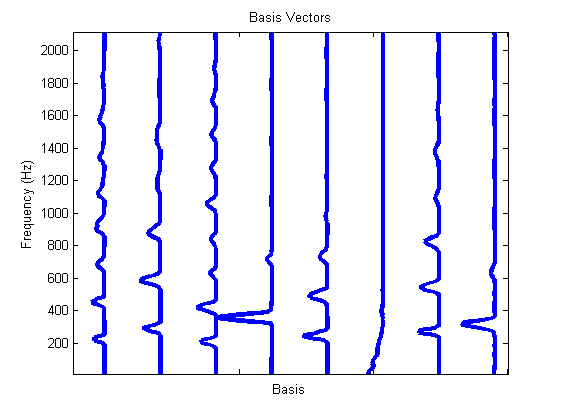}
  \captionof{figure}{Basis Vectors for machali}\label{basis_machami_nmf}
  \label{fig:test2}
\end{minipage}
\end{figure}

When reconstructing the speech audio  using NMF, although the error, $D$(KL divergence) is approximately equal, the audio reconstruction quality has several artifacts which is apparent on listening to the reconstructed audio\cite{A}. \\
Also, by looking at the basis vectors and activations, it is not possible to do an an intuitive analysis of the structure of the audio, as was done in the case of pure instrumental sounds in \emph{Mary}, partly because the activation patterns are much more complex and at many time instants, multiple basis vectors are activated.
The basis vectors also themselves do not reveal much about the speech spectrogram , and there is no apparent pattern in the activations that would indicate temporal patterns in the speech signal.
This calls for an extension to the conventional NMF by introducing a convolutive NMF model\cite{paris}\cite{grady}. Here, instead of each basis vector being a single spectrum, it is a series of consecutive spectra. The $W$ matrix, instead of being a 2-D matrix is a tensor of rank 3 of dimension $M \times R \times T$. The reconstruction in terms of W and H is now given as - 
\begin{align}
V \approx\sum_0^{T-1} W^T\stackrel{t\rightarrow}{H} \label{convnmf_eq}
\end{align}

The $t \rightarrow$ is the shift operator which represents shifting of the columns to the right by t steps. The updated multiplicative updates are as  given in \cite{paris}. T represents the length of each basis \emph{pattern}  (instead of basis \emph{vector}). By adding a temporal dimension to the basis vectors, they have been more expressive in terms of representing patterns in the spectrogram and help capture trends in the temporal dimension which were obscured in the activation vectors in the conventional NMF approach.

This convolutive version of the NMF was applied on the same audio, \emph{Machali}, and the results are shown in Figures 2.9-2.12 . They represent the spectrogram of the original audio, the reconstructed spectrogram, the activation matrix and the basis vectors respectively. All the parameters remained the same (Refer Figure 2.10) and the value of $T$ i.e. the length of each basis pattern was chosen to be 8 samples in the spectrogram domain, which, given the STFT parameters, translates to almost 0.17 seconds in the time domain. Although such a basis length is not entirely reasonable since speech is stationary only upto time segments of around 40ms, the choice is deliberately made for demonstration purposes.
\\
\begin{figure}
\centering
\includegraphics[scale=0.6]{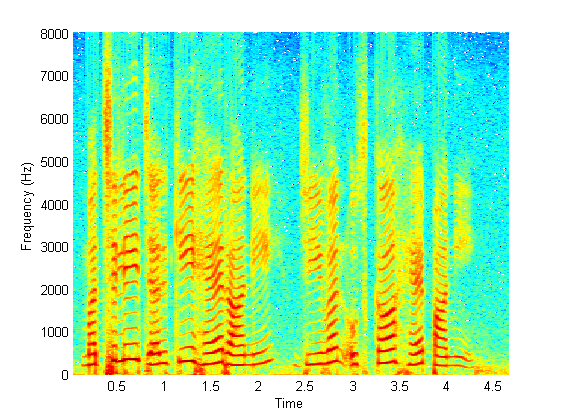}
\caption{Spectrogram of \emph{M}. Key STFT parameters - Sampling frequency : $16000kHz$. Window : hanning window of length 1024 samples. Overlap : 756 samples. Size of fft: 1024 }
\end{figure}

\begin{figure}
\centering
\includegraphics[scale=0.6]{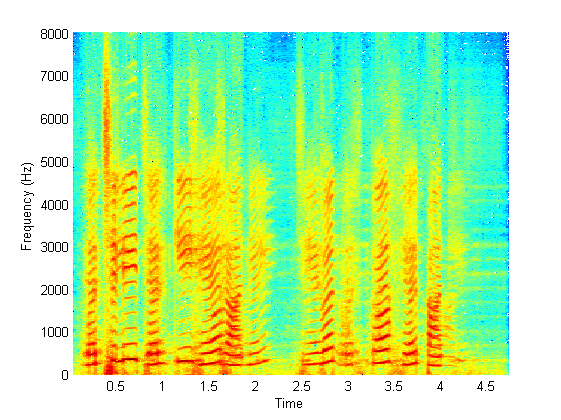}
\caption{Reconstructed Spectrogram of \emph{Machali}. Key STFT parameters - R : 8, T = 8.Sampling frequency : $16000kHz$. Window : hanning window of length 1024 samples. Overlap : 756 samples. Size of fft: 1024 }
\end{figure}

\begin{figure}
\centering
\begin{minipage}{.5\textwidth}
  \centering
  \includegraphics[width=.9\linewidth]{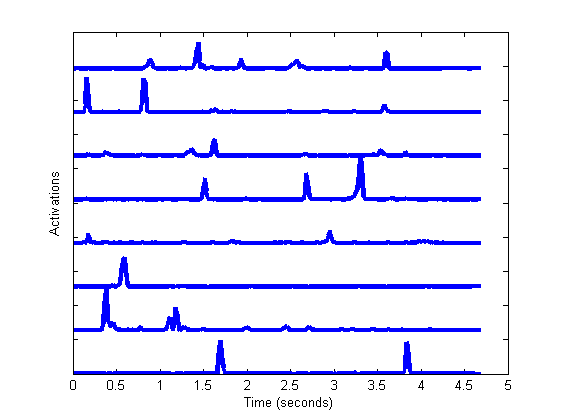}
  \captionof{figure}{Activations for Machali}
  \label{fig:test1}
\end{minipage}%
\begin{minipage}{.5\textwidth}
  \centering
  \includegraphics[width=.9\linewidth]{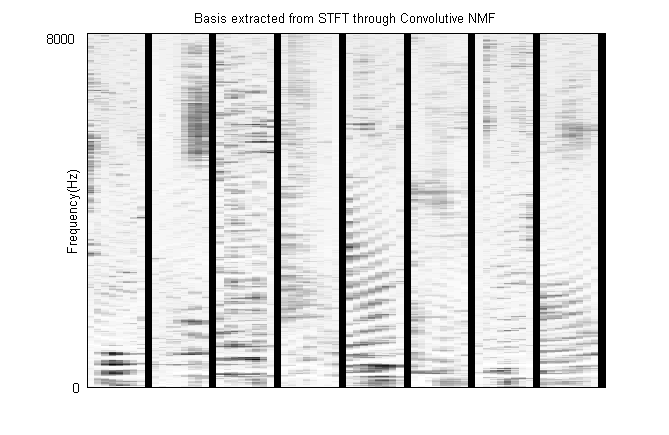}
  \captionof{figure}{Basis patterns for Machali}\label{convolutive_basis_machali}
  \label{fig:test2}
\end{minipage}
\end{figure}

The basis extracted using convolutive NMF are shown in Figure \ref{convolutive_basis_machali}. It is evident that the basis patterns say a lot more about the audio than their counterparts in in conventional NMF. Some of the bases look like spectrograms of speech phones in that they represent the harmonics with various pitch inflections. An audio reconstruction confirms that some basis patterns do represent specific speech phones\cite{A} Also, since the activation matrix is much more sparse in this case, when compared to that of the conventional NMF (Figure 2.7). This enables us to pinpoint the exact contribution of each basis to the spectrogram.
The convolutive model has more effectively captured the structure of speech and has provided a much more revealing set of basis objects. The performance of this technique relies heavily on the choice of the two key parameters - the number of basis($R$) and the time-length of each basis($T$). Not much work has been done regarding the choice of these parameters for a given signal. 
While it is intuitive that having more basis will always increase the accuracy of reconstruction, it also increases the time complexity of the decomposition and the marginal gain in accuracy may not be worth the complexity after a certain point.
Although the convolutivs NMF model does not do as well as NMF as far as reconstruction error is concerned, it has been employed in source separation \cite{paris} and dereverberation \cite{baby}. Thus, this representation for speech can be helpful in a few applications.
The next chapter will describe a popular approach for dereverberation through factorizing the magnitude spectrogram.


\chapter{Dereverberation using NMF}

In the time domain, reverberation is modeled simply as the convolution between the source signal and the room impulse response. 
\begin{align}
y(t) = s(t)*h(t) \label{eq :1}
\end{align}
Here, $s(t)$ is the original speech signal, $h(t)$ is the impulse response of the room, henceforth mentioned as the RIR(Room Impulse Response) and $y(t)$ is the observed speech signal. It is assumed that the room impulse response $h(t)$ does not change with time. This implies that the system modeled is a linear time-invariant (LTI) system. In the time domain, inverse filtering techniques have been developed to recover clean speech from reverberant speech through application of an adequate adaptive filter. The objective is to calculate $h(t)$, the room impulse response and apply its inverse filter to $y(t)$ to obtain $s(t)$. One major issue with time domain filtering is its lack of robustness against speaker movements, which cause a change in the phase of the signal, changing the RIR, i.e. $h(t)$. 
This is why, we have chosen the magnitude spectrogram domain to model and remove reverberant effects from speech. That is, given a spectrogram corresponding to a speechs signal which is recorded in a reverberant environment, the objective will be to counter the effects of reverberation and estimate the magnitude of the clean speech signal. As discussed in Chapter 1, there is an evident smearing effect in the spectral feature domain because of reverberation. This is exactly the intuition for the class of techniques that are discussed henceforth.

 \section{Non negative matrix factor deconvolution (NMFD)}
Consider a signal $y(t)$ which is a recorded version of a source signal $s(t)$ in a reverberant environment.  $Y$ is the magnitude spectrogram, obtained by taking the short time fourier transform (STFT) of $y(t)$. Given $Y$, the task is to estimate $X$, which is the magnitude spectrogram of the clean speech signal. A basic model for the relating these two magnitude spectrograms is \cite{naka} \cite{moham}-
\begin{align} Y_k[n] \approx Y_k[n]' = \sum_{\tau = 0}^{\tau = L-1} X_k[n - \tau] * H_k[\tau] \label{convolution}
 \end{align}

Here, $Y_k$ is the $kth$ frequency sub-band of $Y$, similarly, $X_k$ is the $kth$ frequency sub-band of $X$ and $H_k$ is a sub-band filter, of length $L$ which models the smearing of the clean spectrum sub-band. It is $H_k$ and $X_k$ that have to be estimated from the given $Y_k$, under reasonable constraints. Although in much of the literature, $H$, is said to be the STFT of the room impulse response $h[n]$, there is little evidence in literature that it has any obvious relation to the actual room impulse response . For now though, it will be called the RIR and its relationship with the actual room impulse response will be discussed later. $\hat{Y}$, which is the approximation of the magnitude spectrogram of reverberant speech, $Y$, can have an infinite number of decompositions, $X$ and $H$ which satisfy the above equation. Not all of the solutions though, will give us an estimate of the clean speech magnitude spectrogram and the RIR respectively. Thus, constraints have to be used to narrow the solutions to get more meaningful results. More specifically, we need to ensure that $X$ resembles a speech spectrogram and $H$, a room impulse response spectrogram. 

It has been established that speech magnitude spectrograms are sparse across speakers and utterances. This property has been used in \cite{naka} \cite{moham}\cite{gamma} to arrive at a cost function that gives a quantitative measure of the distance from a desirable solution. The cost function is defined as- 
\begin{align} f(X,H) = \sum_{k,t} (Y_k[t] - Y'_k[t])^2 + 2 \lambda \sum_{k,t} |X_k[t]| \label{cost}
\end{align}
The first term signifies the squared sum of the distances of the coefficients of $Y$ and $Y'$ and the second term enforces sparsity of the clean speech spectrogram.Fpr the underlying assumptions and derivation for this cost function, see \cite{naka}. Clearly, a closed-form solution to the above equation is not possible for (3.3) This problem that is similar to the problem of Non-negative Matrix Decomposition(NMF) discussed in Chapter 2, where a similar cost function was used. To minimize the cost function, we used the gradient descent algorithm to arrive at iterative multiplicative updates which converged to a solution. Similarly, for the cost function in (3.3), multiplicative updates can be derived for $X$ and $H$ as well. With the sparsity constraint on $X$, the iterative multiplicatssive updates for the dereverberation problem are given in \cite{naka} as 
\begin{align}X_k[\tau] = \hat{X}_k[\tau]\frac{\sum_{t}\hat{H_k}[t-\tau]Y_k[t]}{\sum_{t}\hat{H_k}[t-\tau]\hat{Y'_k}[t] + \lambda} \label{X update}
\end{align}
\begin{align} H_k[\tau] = \hat{H}_k[\tau]\frac{\sum_{t}\hat{X_k}[t-\tau]Y_k[t]}{\sum_{t}\hat{X_k}[t-\tau]\hat{Y'_k}[t]} \label{H update}
\end{align}
Here, $\hat{H}$, $\hat{Y}$ and $\hat{X}$ are the estimates from the previous iteration. These updates are based on the gradient descent algorithm that converges to a local minimum. No specific property for H is exploited, to avoid scale ambiguity, the sum of the coefficients of each sub-band is scaled down to sum to one i.e. $\sum_{t}H_k[t] = 1 \ \ \forall k$. This constraint is not imposed in the updates. Rather after each iteration, each row of H is normalized to account for this constraint. For speech, the only property that has been used to enforce a constraint on $X$, is its sparsity in the magnitude spectrogram domain. In the next few sections, we modify the above approach to incorporate specific properties.

\section{Activation Deconvolution}
The technique of non-negative matrix factorization efficiently separates the temporal and the spectral features of the magnitude spectrogram of the speech signals in the activation matrix and the basis vectors respectively. Reverberation can be modeled as a largely temporal phenomenon where delayed and attenuated versions of the signal corrupt the recorded signal. Our hypothesis is that since reverberation is a temporal phenomenon, its effects will be localized to the temporal part of the non-negative factors, which is the activation matrix

\begin{figure}
\centering
\includegraphics[scale=0.6]{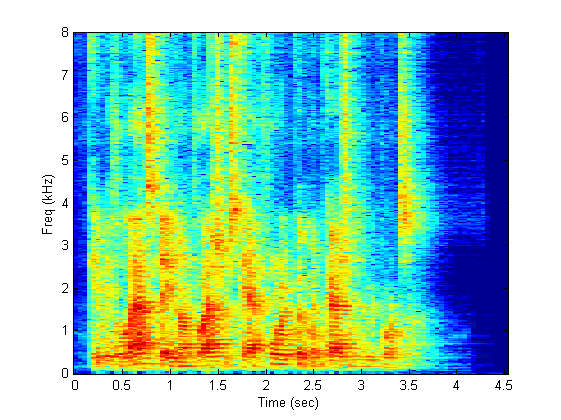}
\caption{ Spectrogram for a reverberated speech signal. Key parameters - Sampling frequency : $16000kHz$. Window : hanning window of length 1024 samples. Overlap : 512 samples. Size of fft: 1024 } \label{input_spec}
\end{figure}

\begin{figure}
\centering
\includegraphics[scale=0.6]{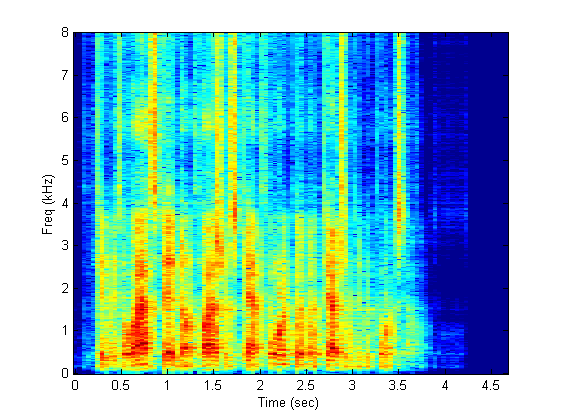}
\caption{Spectrogram after the activations were processed. Notice the reduction of the spectrogram smearing from Figure \ref{input_spec} . Key parameters - Sampling frequency : $16000kHz$. Window : hanning window of length 1024 samples. Overlap : 512 samples. Size of fft: 1024 } \label{filtered spec}
\end{figure}

\begin{figure}
\centering
\begin{minipage}{.5\textwidth}
  \centering
  \includegraphics[width=.9\linewidth]{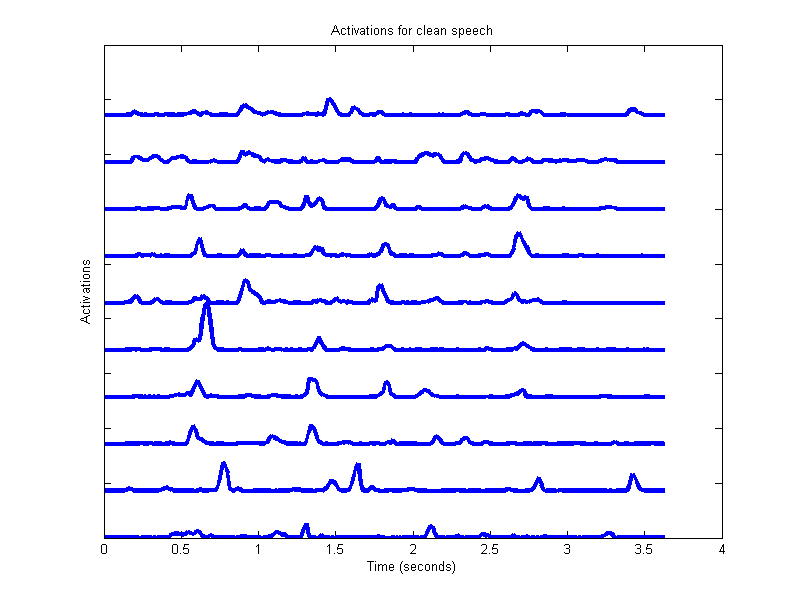}
  \captionof{figure}{Activations in clean speech} \label{activations_clean}
  \label{fig:test1}
\end{minipage}%
\begin{minipage}{.5\textwidth}
  \centering
  \includegraphics[width=.9\linewidth]{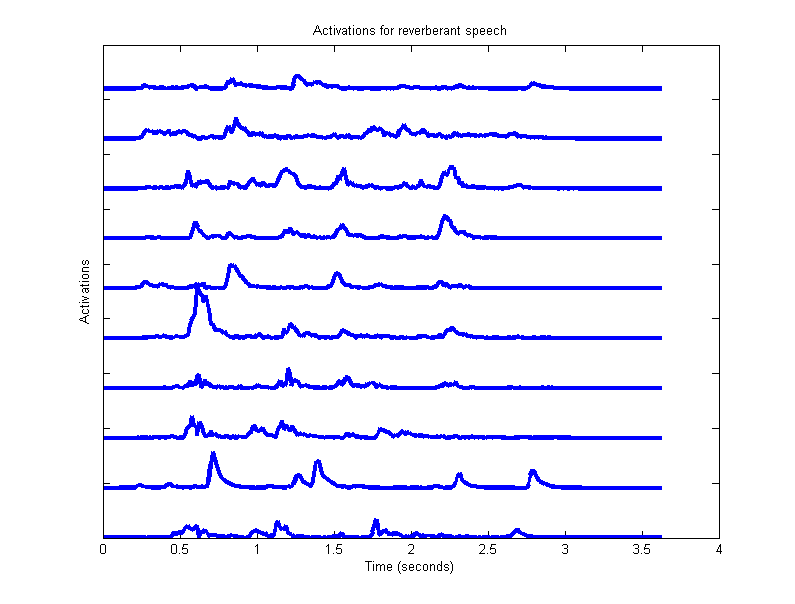}
  \captionof{figure}{Activations in reverberant environment} \label{activation_reverb}
  \label{fig:test2}
\end{minipage}

\end{figure}

Figures \ref{activations_clean} and \ref{activation_reverb} are the activation patterns for two speech signals.Figure \ref{activations_clean} shows the activation pattern that was obtained by applying NMF on a sentence recording from the TIMIT database. Figure \ref{activation_reverb} on the other hand shows activation patterns for the same audio signal, with reverberation added. There is a strong correlation between the activation patterns. We hypothesis that an activation vector for the clean speech spectrogram, $A_i$, corresponding to a basis vector $i$ undergoes the following transformation when reverberation is added to  the clean speech signal.
\begin{align}
A_i^{reverb}(t) = A_i(t)*h(\tau)
\end{align}
Therefore, given the reverberant signal, we can obtain the activation vectors $A_{i}^{reverb}$ . Then using the Non-Negative factor deconvolution technique, presented in the previous section, $A_i$ can be isolated. Notice how this equation is similar to Equation \ref{convolution}. Thus, we use the same updates in Equations \ref{X update} and \ref{H update} to estimate the clean speech activation vector $A_i(t)$ from the reverberant activation vectors $A_i^{reverb}$. A spectrogram is then reconstructed using these activations. Figure \ref{input_spec} shows the spectrogram of a speech signal from the TIMIT database with added reverberation. Figure \ref{filtered spec} shows the spectrogram obtained after the activations were processed using the above mentioned algorithm. It can be clearly seen that the smearing of the spectrogram has been visibly reduced. 

\section{NMFD using NMF model for speech}
The imposition of constraints in NMFD ensures that the solution converges to a clean speech spectrogram. A better model for speech or RIR($X$ and $H$ respectively) will ensure convergence to a better solution. In section 3.1, we have only exploited that speech is sparse in the magnitude spectrogram domain. NMF has been shown to be an efficient way to represent speech by decomposing it into factors with significantly less dimensions. It is based on the assumption that speech spectrograms are low rank. NMF tries to find an encapsulating basis vector set to represent the whole spectrogram.  This representation has been shown to work in separation  \cite{paris} and dereverberation \cite{baby} very well. We now incorporate this representation into the Non-Negative factor deconvolution model discussed in the previous section as suggested in \cite{moham}.

Given input magnitude spectrogram, $Y$, of a time domain reverberated signal, $y(t)$ we can use NMF to decompose into a basis matrix $W$ and an activation matrix $X$ such that $ S = W*X$ and $Y \approx S$. 
Now, we will introduce a cost function, as given in \cite{moham}, which quantifies the distance of the current estimate from the desired solution. This cost function is based on the Kullback-Leiber(KL) divergence, which has been shown to be better suited to speech applications than the sum squared error used in \cite{naka}. The cost function using KL divergence, without including the NMF representation for speech is given by -
\begin{align}
f(Y,S,H) = \sum_{k,t} KL( Y(k,t) \ | \ \sum_{\tau = 0}^{L}H(k, \tau)S(k,t-\tau) ) + \lambda \sum_{k,t}S(k,t)
\end{align}

$L$ is the length of the RIR in the spectrogram domain. $\lambda$ is the sparsity parameter. Now, NMF models $S$ as a combination of two matrices, $W$ and $X$. The modified cost function, incorporating this relationship is -
\begin{align}
f(Y,W,X,H) = \sum_{k,t} KL(Y(k,t)|\sum_{\tau = 0}^{L}H(k, \tau)\sum_{r =1}^{R}W(k,r)X(r,t-\tau)) + \lambda \sum_{r,t}X(r,t)
\end{align}
For this modified cost function, the multiplicative updates for the gradient descent algorithm are given in \cite{moham}.
\begin{align}H(k,\tau) = \hat{H}(k,\tau) \frac{\sum_{t}Y(k,t)\hat{S}(k,t - \tau)/\hat{Y'}(k,t)}{\sum_{t}\hat{S}(k,t - \tau)}
\end{align}

\begin{align}W(k,r) = \hat{W}(k,r) \frac{\sum_{t,\tau}Y(k,t)H(k,\tau)\hat{X}(r,t - \tau)/\hat{Y'}(k,t)}{\sum_{t,\tau}H(k,\tau)\hat{X}(r,t - \tau)}
\end{align}
\begin{align}X(r,t) = \hat{X}(r,t)\frac{\sum_{t,\tau}Y(k,t + \tau)H(k,\tau)W(k,r)/\hat{Y'}(k,t + \tau)}{\sum_{k,\tau}H(k,\tau)W(k,r) + \lambda}
\end{align}

Here,
$Y'[k,t] = \sum_{\tau}H[k,t]*S[k,t - \tau]$  and $S = W*X$. Also, $\hat{Y'}, \hat{X},\hat{S} \ and \  \hat{H}$ correspond to the values of $Y,X,S \ and \ H$ from the previous iteration respectively. 
After an appropriate number of iterations, an estimate for the clean speech spectrogram is given as $S = W*X$. 
Including the NMF representation into the NMFD approach improves the quality of the clean speech spectrogram, $S$ \cite{moham}. This has been shown in literature and will be demonstrated experimentally in Chapter 4. In the next section, we incorporate another fundamental property of speech, namely, temporal continuity. 
\section{Exploiting temporal dependencies of speech }
In section 3.2 and 3.3, two properties of speech have been incorporated into the dereverberation model, the sparsity of magnitude spectrogramdomain and the low rank nature of the magnitude spectrogram (through NMF decomposition). Temporal continuity is another property of speech that can possibly be exploited to model the speech signal better. In this section, two techniques will be discussed which model the temporal dynamics of speech and can be implemented in a similar fashion to the previous techniques with modifications to the update rules.
\subsection{Frame stacking} 
A frame stacked approach proposed in \cite{moham} is taken to exploit the fact that the magnitude spectrograms is temporally dependent. A sliding window of length $T_{stack}$ is passed through the time axis of $Y$ and each resultant frame is stacked to form a vector of dimension $K \times T_{stack}$. These vectors are then put together to form a higher dimensional matrix, $\tilde{Y}$. Now, $\tilde{Y}$ is decomposed using NMF and the algorithm follows the same as in the previous section. $W$ will now be a $KT_{stack} \times R$ matrix. The multiplicative updates for $W$ and $X$ remain the same as in (3.10) and (3.11), and the modified updates for $H$ as given in \cite{moham} are -
\begin{align}H(k,\tau) = \hat{H}(k, \tau)\frac{\sum_{l=1}^{T_{stack}}\sum_{t}\tilde{Y}(f_l,t)S(f_l,t -\tau)/\hat{Y'}(f_l,t)}{\sum_{l=1}^{T_{stack}}\sum_{t}S(f_l,t - \tau)}
\end{align}
Here, $f_l = k + l(K-1)$ . This update is essentially taking an average over all the stacked up spectrograms for an estimate of $H$. After convergence, the clean speech magnitude spectrogram is estimated as $S=G*Y$ where G is the gain function given by -
\begin{align}
G(k,t) = \frac{\sum_{l=1}^{T_{stack}}\sum_{r}W_{final}(k_l,r)X_{final}(r,t)}{\sum_{l=1}^{T_{stack}}\sum_{t}H_{stack}(f_l,\tau)W_{final}(k_l,r)X_{final}(r,t- \tau)}
\end{align}
The subscript final denotes that the values used are after all updates have been done and the algorithm has converged. $H_{stack}$ is the RIR matrix. The results and relative performance for this technique is discussed in Chapter 4. As per our experiments, it does not improve the performance of the algorithm discussed in the previous section. In the next section, we introduce another technique through which the temporal continuity of speech can be exploited.

\subsection{Convolutive NMF}
Another way to exploit the continuous nature of speech signals is by using the convolutive Non-negative matrix factorization . This technique was discussed in Section 2.3 and is used to extend the the expressive power of the basis vectors which constitute $W$ by adding to them a temporal dimension. It has already been used in source separation and enhancement applications. \cite{baby} uses the Convolutive model for dereverberation, but relies on learning the basis offline.
To include this model for speech in the dereverberation algorithm, the modified update equations are given as -

\begin{align}X(r,t) = \hat{X}(r,t)\frac{\sum_{\alpha = 1}^{T_{base}}\sum_{k,\tau}Y(k,t + \tau)H(k,\tau)W(k,r,\alpha)/\hat{Y'}(k,t + \tau)}{\sum_{k,\tau}H(k,\tau)W(k,r) + \lambda}
\end{align}

The update equations for $W$ and $H$ remain the same. Here, $T_{base}$ is the time length of the exemplars which comprise $W$.  These updates were adopted from \cite{baby} where they have been used to achieve dereverberation is a supervised manner. 

In this chapter, we have discussed four techniques to handle reverberation in the magnitude spectrogram domain. Non-Negative Factor deconvolution(NMFD) forms the core of all these techniques. Activation matrix deconvolution is essentially NMFD applied on the activation matrix instead of the speech spectrogram. Further, NMF representation for speech is included in the NMFD technique. Then, to exploit the temporal dependencies of speech, the frame stacked model and the convolutive NMF representation  were included. All these techniques are implemented in very similar ways. Each of them has an associated cost function which is to be minimized to get non-negative matrix factors with desirable properties. Using the updates suggested by \cite{lee}, each of these cost functions are minimized using multiplicative iterative updates.

In the next chapter the implementation details are presented and performances  of these techniques are compared based on two objective measures - Perceptual evaluation of speech quality (PESQ) and Cepstral Distortion(CD).

\chapter{Experiments and Results}

\ifpdf
    \graphicspath{{Chapter3/Figs/Raster/}{Chapter3/Figs/PDF/}{Chapter3/Figs/}}
\else
    \graphicspath{{Chapter3/Figs/Vector/}{Chapter3/Figs/}}
\fi
This chapter discusses the experiments done to evaluate the techniques described in the previous chapter. The speech samples used were taken from the TIMIT database \cite{TIMIT}. 40 different sentences were chosen from the database for this purpose. These were each sampled at 16kHz. To add reverberation to these clean speech signals, room impulse responses provided by the REVERB 2014 challenge were used. In particular 3 room impulse responses from 3 rooms each with reverberation times (t60) 0.25s, 0.5s and 0.7s were used. Although these responses were recorded using multiple channels, only one channel per room impulse response was used here since these techniques apply to single channel dereverberation.   Reverberant speech signals we generated by convolving clean speech signals with these room impulse responses. MATLAB was used to implement all the techniques. The code for the NMFD algorithm discussed in Section 3.1 was otained from \cite{gamma}. All other techniques have been implemented by us. 
For converting time domain signals into magnitude spectrograms, a sliding square root hann window \cite{moham} length 64ms  was used with an overlap of 16ms between frames.

All the techniques will be quantitatively compared on their performances in dereverberating speech recordings. The techniques are applied to the magnitude spectrogram of the reverberant speech input to obtain a clean magnitude spectrogram. This spectrogram, using the phase information of the input signal is then converted to the time domain. 
To quantify the performance, the following key objective measures are used - 
\begin{enumerate}
\item \textbf{Perceptual evaluation of speech quality (PESQ)} \\
PESQ is a widely used measure to judge the quality of speech as heard by humans. It uses the clean speech as a reference to score the output speech on its quality. A higher score indicates better quality of speech. The performance of a dereverberation technique is good if the PESQ score of the output is more than that of the input.
\item \textbf{Cepstral distortion(CD)} \\
Mel-cepstral coefficients are perhaps the most frequently used features for speech recognition. A distortion in the cepstral coefficients can thus significantly decrease performance in recognition. Therefore, an important measure of how much a speech signal has been affected by reverberation is the amount of distortion introduced in the cepstral coefficients. A quantitative measure of this is the Cepstral distortion. 

\end{enumerate}

\section{Dereverberation using NMFD}

For this technique, discussed in Section 3.1, the multiplicative updates according to (3.4) and (3.5) were carried out for 20 iterations. $\lambda$ , the sparsity parameter was fixed to be $\sum_{k,t}Y \times 10^{-8}$ as suggested in \cite{naka}. However experimentally, we found that having a sparsity constraint did not help in improving algorithm performance.
Initially, the clean speech estimate, $X$ was set to be equal to $Y$ and $H$ was initialized as a linearly decaying filter.
The length of the filter, was fixed at $L=11$ frames in the spectrogram, i.e. a time domain span of 352ms. The average improvements in the PESQ scores and the CD scores are shown in Figures 4.10 and 4.11. Figures 4.2, 4.3 and 4.4 shows the spectrogram of the reverberant speech, the output clean speech and the output filter, $H$. In the output spectrogram, smearing of the features is reduced, specially in the sparser regions.
\begin{figure}
\centering
\includegraphics[scale=0.8]{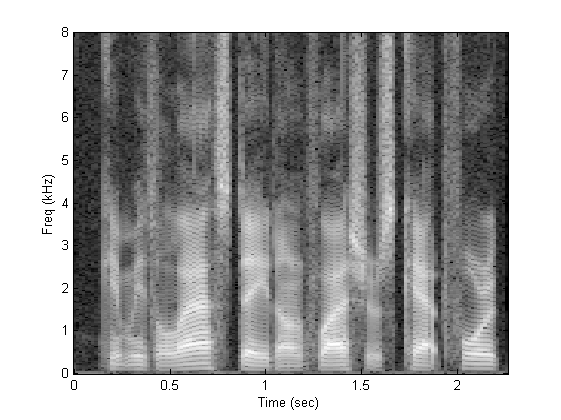}
\caption{Clean speech Spectrogram }
\end{figure}

\begin{figure}
\centering
\includegraphics[scale=0.8]{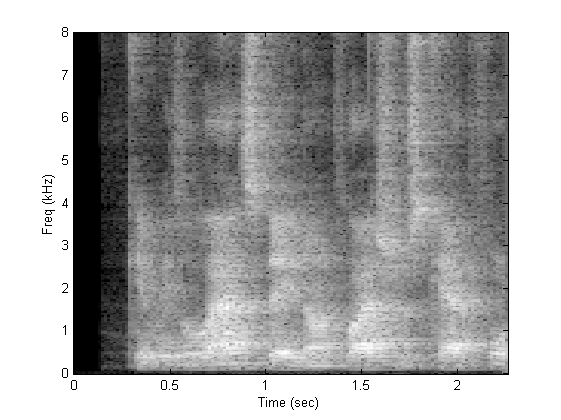}
\caption{Reverberated Spectrogram }
\end{figure}
\begin{figure}
\centering
\includegraphics[scale=0.8]{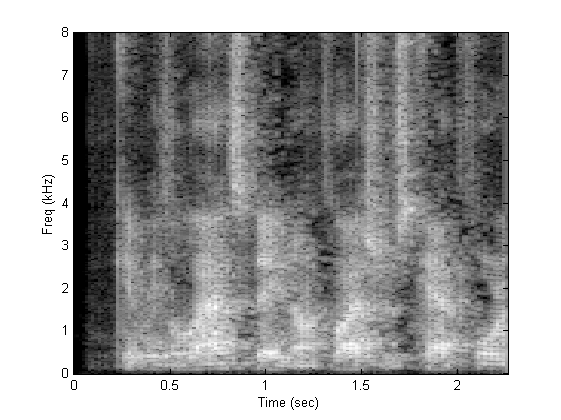}
\caption{Dereverberated Spectrogram }
\end{figure}
\begin{figure}
\centering
\includegraphics[scale=0.8]{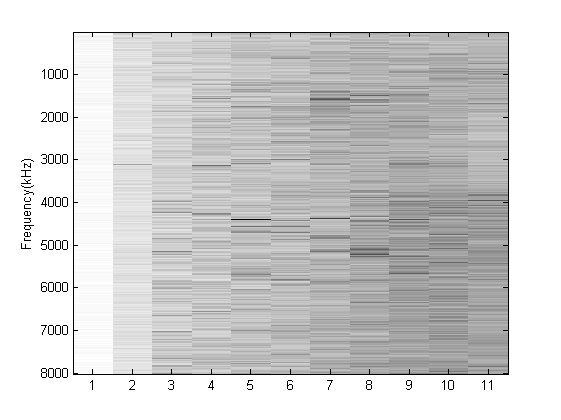}
\caption{Recovered H filter. the filter length was chosen to be 11 frames. The sum of each of the rows of this H matrix is 1.  The X axis represents the frame numbers, where each frame is temporally equivalent to 32ms.}
\end{figure}
The recovered H filter, in Figure 4.4, is displayed using a decibel scale to make it seem more uniform. The initial values of the sub-band filters are much larger than the later values by an order of magnitude. To observe the variation of the performance of the algorithm with varying lengths for the filter, H for each RIR, the length of the filter was varied from 128ms to 640ms. The results are shown in Figure 4.5

\begin{figure}
\centering
\includegraphics[scale=0.6]{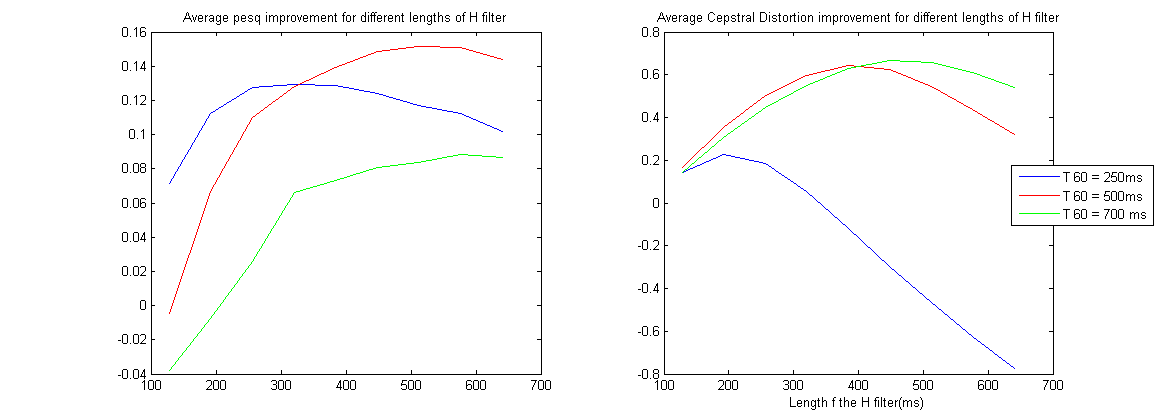}
\caption{Varying the filter length for different T60. Higher T60 values had a higher performance for longer lengths of the H filter}
\end{figure}
It is clear from the results in Figure 4.5, that the length of the filter H, has some correlation to the degree of reverberation in the recorded speech signals. For a higher T 60, the PESQ and Cepstral Distortion are best for greater lengths of $H$ when the T 60 is higher and for lower lengths of $H$ for lower T 60 values. This makes intuitive sense since a larger T 60 value implies that the effect of reverberation on a frame affects frames further in the magnitude spectrogram than it would if the T60 was lower.
\section{Activation Deconvolution}
\begin{figure}
\centering
\includegraphics[scale=0.6]{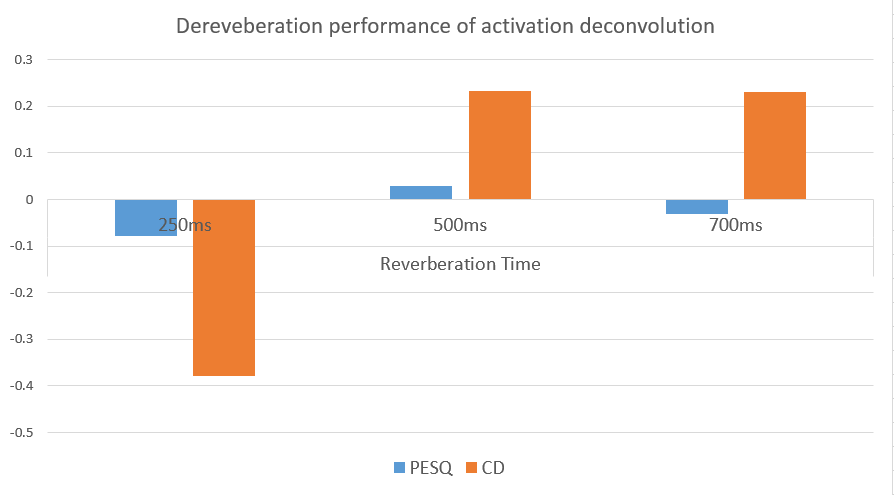}
\caption{Results for the Activation deconvolution technique. The improvements in PESQ and CD scores for different levels of reverberation are shown. Note that a negative value indicates that the output is worse than the reverberant input for the particular measure}\label{act_decon_results}
\end{figure}

The activation deconvolution technique was applied over the same 40 sentences from TIMIT database using the same updates used in (3.4) and (3.5) for NMFD. The number of basis chosen was $R = 20$ and the length of the filter h(t) in Equation 3.6 was chosen to be $L=11$ frames, which translates to 352ms. The output speech was measured for improvement in PESQ, Cepstral Distortion and SRMR scores. The mean results are shown in Figure \ref{act_decon_results}. \\
It can be seen that the cepstral distortion scores have improved for T60s 500ms and 700ms. PESQ score has only increased for T60 500ms. There are audible artifacts in the output sound as well. We infer that the effectof reverberation on the activation vectors has not been modelled appropriately through this approach. A better model for this is needed.

\section{NMFD using NMF model for speech}
Including the NMF model for speech, the updates given by (3.8), (3.9) and (3.10), were carried out for 20 iterations. The number of basis vectors used to represent the speech signals, $R$, used was 10. Having more number for basis vectors gave very marginal increase in performance while adding to the computation time. This is clear from the increase in the mean PESQ and CD performance improvement across the 40 senteneces from TIMIT, shown in Figures 4.7 and 4.8. The sparsity parameter, $\lambda$ was the same as in the previous section. The result for this algorithm are shown in Figure 4.9, for one sentence from the TIMIT database.

\begin{figure}
\centering
\includegraphics[scale=0.8]{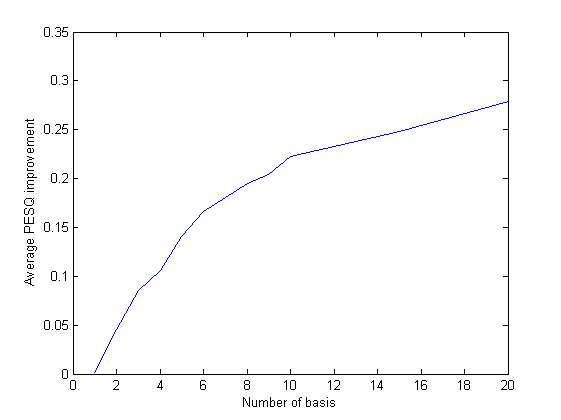}
\caption{Number of basis on performance}
\end{figure}

\begin{figure}
\centering
\includegraphics[scale=0.8]{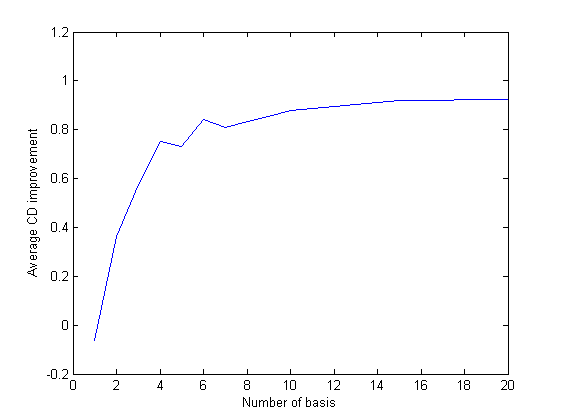}
\caption{Number of basis on performance. Both figure 4.7 and 4.8 unequivocally suggest that more basis vectors in the NMF decomposition will provide better results for dereverberation. However, the marginal increase in performance after a certain number of basis is minimal}
\end{figure}
\begin{figure}
\centering
\includegraphics[scale=0.8]{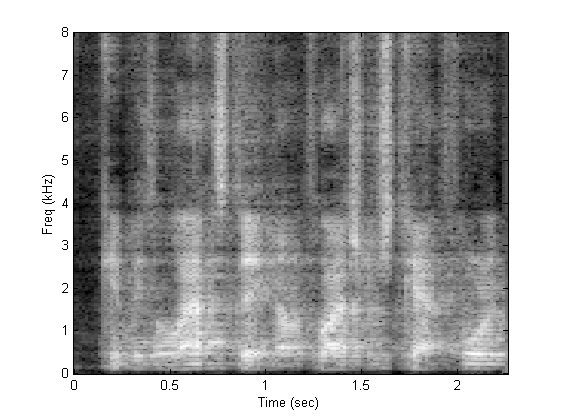}
\caption{Output spectrogram for NMFD with NMF model}
\end{figure}

A comparative analysis is done, and the results reproduced in Figures 4.10 and 4.11 to show the improvements gained in the PESQ scores and Cepstral Distortion Scores by introducing the NMF model for speech. It can be seen that significant improvement in PESQ scores was observed. The improvement in the Cepstral distortion scores was not significant.
\begin{figure}
\centering
\includegraphics[scale=0.6]{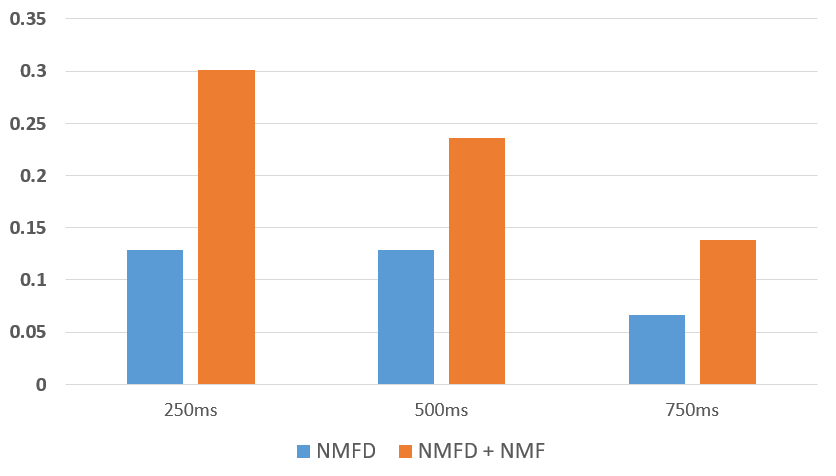}
\caption{Output for NMFD with NMF model}
\end{figure}
\begin{figure}
\centering
\includegraphics[scale=0.6]{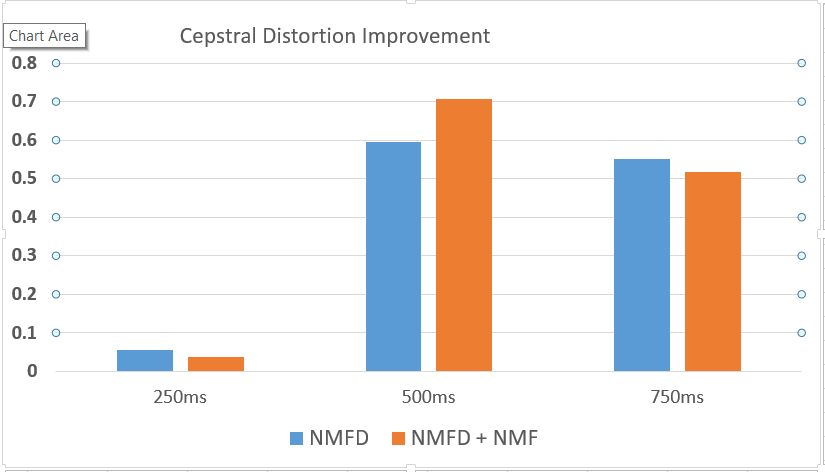}
\caption{Output for NMFD with NMF model}
\end{figure}

\section{Frame stacked model}
The frame stacked model discussed in Section 3.4.1, aims to exploit the temporal continuity of speech through stacking the magnitude domain spectrogram in a sequential manner. The updates for this model are as given in (3.12). The temporal context to use for stacking, quantified by the variable, $T_{stack}$ . These updates are given 20 iteration to converge and then, the gain function, $G$ is calculated using (3.13) is calculated. 

The results for the stacked model were not found to be promising. In fact, performance gets worse even if $T_{stack} = 2$. It does not outperform the simple NMF model for any of the reverberant scenarios, as far as improving the PESQ and the Cepstral Distortion scores are concerned. This agrees with the results in literature where this model has not been able to outperform the simple NMF model for speech.

\section{NMFD and the Convolutive NMF model }
This section discusses the results of NMFD with the Convolutive NMF model used for speech. The modified updates for this approach are given in (3.14). This approach has not yet been presented in any literature to the best of our knowledge, although \cite{baby} has used the convolutive NMF model as part of a supervised approach. The updates are derived on similar lines as the stacked model. Again, of major importance is the temporal context used. The length of the convolutive basis, $T_{base}$, has a major influence on the algorithms performance. 
\begin{figure}
\centering
\includegraphics[scale=0.6]{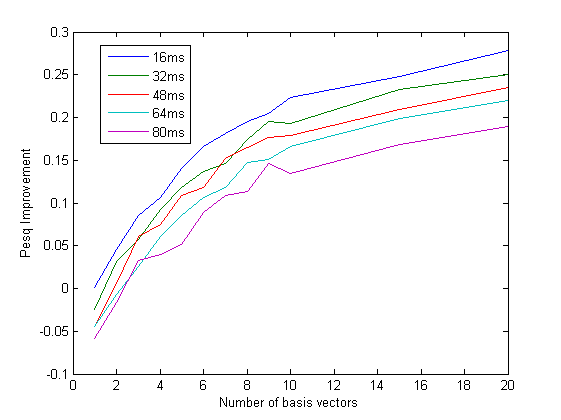}
\caption{PESQ score versus number of bases(x-axis) and length of bases(plot lines).Figures 4.12 and 4.13 show the improvement in performance with increasing number of basis vectors, across different basis lengths. A basis length of 16ms, which is equivalent to one frame, implies a regular NMF. All other multiples of 16ms correspond to convolutive basis vectors. Each plot line corresponds to a fixed length of the convolutive basis}
\end{figure}

\begin{figure}
\centering
\includegraphics[scale=0.6]{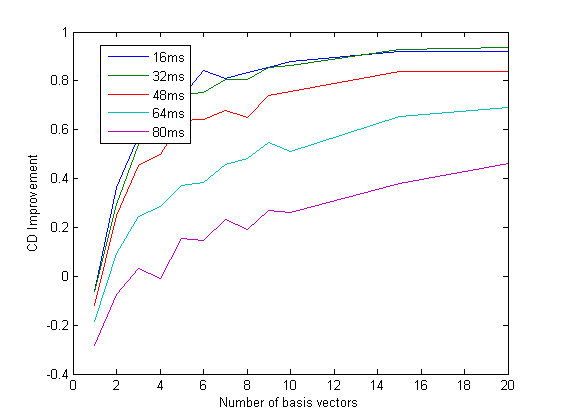}
\caption{CD scores. Figures 4.12 and 4.13 show the improvement in performance with increasing number of basis vectors, across different basis lengths. A basis length of 16ms, which is equivalent to one frame, implies a regular NMF. All other multiples of 16ms correspond to convolutive basis vectors. Each plot line corresponds to a fixed length of the convolutive basis}
\end{figure}

It is clear from Figures 4.12 and 4.13 that increasing the length of the basis, $T_{base}$ has a detrimental impact on the performance of the algorithm, both in terms of PESQ scores as well as Cepstral Distortion scores but more number of bases is beneficial in both measures. Each plot line, corresponding to a longer basis vector is lower than that of one with a shorter one. It can be concluded that Convolutive NMF representation of speech does not help in increasing dereverberation performance.

\section{Discussion}
As per the measures, it is clear that NMFD, with an NMF representation of speech is the best among all techniques for dereverberation. Our hypothesis as to why it is so is that by applying the sparsity constraint on the activations, rather than on the whole spectrogram, as in NMFD, it has provided a better model for speech magnitude spectrograms. Also, activation deconvolution, although does not do as well as NMFD provided an alternative approach to tackle dereverberation in the magnitude spectrogram domain. The approach, as it is now, also introduces artifacts in the output speech. 
As far as exploiting the temporal dynamics of speech is concerned, neither the frame stacked model, nor the convolutive model improve the performance of the algorithm as far as the objective measures are concerned. 
The qualitative results obtained in this chapter corroborate with the existing literature \cite{moham} although the exact improvements in objective measures obtained are still not as high. This might be due to different datasets used or difference in implementation details. The summary of all the results is given in Table 4.1 and 4.2

\begin{table}
\centering
\caption{Summary of the PESQ improvement by discussed techniques}
\label{my-label}
\begin{tabular}{llll}
                             & \multicolumn{3}{c}{\textbf{T60}} \\
\textbf{Technique}           & 250ms     & 500ms     & 750ms    \\
NMFD                         & 0.1291    & 0.1283    & 0.0662   \\
NMFD + NMF                   & 0.301259  & 0.235671  & 0.138228 \\
Stacked model (T\_stack = 3) & 0.135805  & 0.192561  & 0.094765
\end{tabular}
\end{table}

\begin{table}[]
\centering
\caption{Summary of the PESQ improvement by discussed techniques}
\label{my-label}
\begin{tabular}{llll}
                             & \multicolumn{3}{c}{\textbf{T60}} \\
\textbf{Technique}           & 250ms     & 500ms     & 750ms    \\
NMFD                         & 0.0552    & 0.5963    & 0.5516   \\
NMFD + NMF                   & 0.0381    & 0.7068    & 0.5180   \\
Stacked model (T\_stack = 3) & 0.0312    & 0.4021    & 0.4974  
\end{tabular}
\end{table}

\chapter{Conclusions and Future Work}  

\ifpdf
    \graphicspath{{Chapter1/Figs/Raster/}{Chapter1/Figs/PDF/}{Chapter1/Figs/}}
\else
    \graphicspath{{Chapter1/Figs/Vector/}{Chapter1/Figs/}}
\fi

\section{Conclusions}
In this work, NMF based approaches to dereverberation has been studied and tested. An attempt has been made to survey and reproduce results found in literature and tweak the parameters to gain a better understanding of why certain techniques work better. 
First, the NMFD approach, which models reverberation using sub-band filtering works well for dereverberation. The filter, $H$ captures the relation between the source magnitude spectrogram and its reverberated version for each frequency sub-band. 
Also, it is clear that  better model for speech or the room impulse response will definitely help in dereverberation. This is verified by the fact that a better representation for speech, the NMF representation improves the performance of NMFD. 
Convolutive NMF model for speech, although intuitively is a good model for speech, does not improve performance of the dereverberation algorithm. In general, for unsupervised approaches to dereverberation or separation, the convolutive model does not outperform the regular NMF model. It can be concluded that having a more detailed set of basis is only beneficial to capture more information about about the speech and thus is only useful where a supervised approach, using a dictionary of basis is used.
Additionally, different models for the obtained sub-band filter H have been adopted by different authors, but a general consensus on what the room impulse response or H should look like is lacking in literature. Its relation with the magnitude spectrogram of the time domain room impulse response is also unclear and is a possible area to explore. 

\section{Future work}
First, speech recognition rates for the listed approach using a standard recognition system need to be obtained. Since the results obtained do not match quantitative figures quoted in literature and it is difficult to reproduce the experiments exactly, word error rate improvement can provide a benchmark as to where these techniques stand in comparison to state of the art dereverberation techniques. Models for $H$, the set of sub-band envelops, should be explored to understand and improve the assumptions made about it. 
Activation deconvolution technique that was developed does not work as well as techniques in literature, but it does provide an alternative approach to dereverberation. The effects of reverberation on the NMF activation matrix need to be understood and modeled  for this approach to work better.

   .

\appendix 


\backmatter


\newcommand{\acknowledgements}{

\thispagestyle{plain}

\begin{center}{\huge{\textit{Acknowledgements}} \par}\end{center}

\vspace*{15px}


\noindent I would first like to thank the staff, faculty and students of the Electrical Engineering department of IIT Bombay. They have been extremely generous with me. Being a part of my department, I got to meet some of the most hardworking, honest and warm-hearted people I know. Some of that has seeped in I hope. \\
Specifically, I'd like to thank Prof. V.Rajbabu for tolerating my terrible work ethic and being patient with me. I can only imagine how hard it must be to have someone like me working for you. Thank you sir. \\
I would like to thank my friends for not letting me take life seriously and always being around for (mostly)pointless conversations. \\
Finally, I'd like to thank my mother and brother, the two most important people in my life. I am hugely indebted to them for whatever little I have achieved. 

\vspace*{15px}

\begin{flushright}
{Signature: ......................................\\[0.4cm]}

{\textbf{\authorName}\\[0.0cm]\rollNo\\[2.0cm]}
\end{flushright}

\begin{flushleft}
{Date:} 05 \currentmonth { } \currentyear\\
\end{flushleft}
}


\label{Bibliography}
\lhead{\emph{Bibliography}}

\bibliographystyle{unsrtnat} 

\bibliography{Bibliography} 

\clearpage


\clearpage

\addcontentsline{toc}{chapter}{Acknowledgements}
\acknowledgements

\end{document}